# Catalog of spiral arm tangents (Galactic longitudes) in the Milky Way, and the age gradient based on various arm tracers

Jacques P Vallée
Herzberg Astronomy and Astrophysics Research Center, National Research Council of Canada
ADDRESS 5071 West Saanich Road, Victoria, British Columbia, Canada V9E 2E7
ORCID *http://orcid.org/0000-0002-4833-4160*
EMAIL jacques.p.vallee@gmail.com

**Abstract.**

An updated catalog of 205 observed tangents to the spiral arms (in Galactic longitudes) since 1980 is presented. This represents an addition of 80 arm tangents in 6 years (since 2016). Most arm tangents are observed at telescopes in the radio régime.

In this study, the separation of each arm tracer from the dust lane is analysed to obtain the relative speed away from the dust lane (an age gradient). Each arm tracer is observed to be separated from the dust lane, showing an age gradient of about *11.3 ±2 Myr/kpc* across the spiral arm – a relative speed away from the dust lane of about *87 ±10 km/s*.

## 1. Introduction

Our Milky Way galaxy has been observed at optical wavelengths, revealing 4 major spiral arms full of stars and dust. This picture was confirmed also at radio wavelengths for gas and protostars. Our knowledge of the distance from the Sun to the Galactic Center took a while to measure appropriately, assumed to be zero until the work of Shapley around 1910 or so (see Fig. 2 in Vallée 2005), and now being closer to 8 kpc For a short comprehensive review, see Vallée (2017b).

The precise measurements of **the Galactic Longitudes of the arm tangents can reveal new details of the spiral structure** of our Galaxy, thus **:** **-*** One could find if there is a separation of some arm tracers among each others (CO versus dust – see Figure 1 and Table 4 in Vallée 2014a); **-*** One could find if there is a mirror-image of this separation in Galactic Quadrants IV (shock at right) versus I (shock at left) – see Figure 2 in Vallée 2016b; **-*** One could find if there is an age gradient from the shock to the dust lane – see Figure 6 in Vallée 2021b and Figure 2 in Vallée 2022; **-*** One could find an estimate of the angular speed of the density wave spiral pattern (shock front) – see Table 1 in Vallée 2021a; **-*** One could find an estimate and variation of the spiral arm width, from multi-tracers, with increasing Galactic radius – see Figure 2 in Vallée 2020a; **-*** One could find an estimate of the co-rotation radius in the disk of the Milky Way – see Table 2 in Vallée 2018b; **-*** One could find an estimate of the large-scale pitch angle of each spiral arm, using arm tangents from two Galactic quadrants – see Figure 1 in Vallée 2015; **-*** One could find an estimate of the starting point of each spiral arm, near the Galactic nucleus – see Figure 2 in Vallée 2016a. The publication of this updated catalog may spur other researches, and comparisons with other catalogues (Gaia, etc), so that new results may emerge in the future, **-*** while comparisons with predictions from different spiral arm formation theories (Vallée 2022) may help eliminate some theories or spur new theoretical developments.

Historically, the first telescope scans along the Galactic longitudes was done at optical wavelengths, then at radio wavelengths and at infrared wavelengths – see Section 2.1 below.

Different methods have been attempted to measure the arm tangents, notably for stars, protostars (radio masers), dust at different near-infrared, mid-infrared, and long-infrared wavelengths, HI atom and CO gas spectroscopy - see Section 2.2 below.

A considerable number of arm tangents (a line from the Sun tangent to a spiral arm in the Galactic disk) have been measured and published in the literature; we selected all those published since 1980 with a high accuracy. Scanning along the Galactic disk, telescopes scan in Galactic longitude using a known tracer (CO, HI, dust, star count, etc) and can thus record the precise Galactic longitude when a peak intensity is observed in that tracer. This peak occurs when the telescope looks along a long distance inside the arm, corresponding to the line-of-sight being tangential to the spiral arms. Published observations of telescope scans, in Galactic longitude, have shown a peak intensity when being tangents to a spiral arm, as seen from the Sun's position. The scans show a consistent value, when taken with similar but individual telescopes, provided that the angular telescope beam is similar from one telescope to the next – in order to collect emission from the same physical properties of the arm tracer under study.

The arrival of new observations of the arm tangents, in new arm tracers, in the literature can help find more precise locations, and thus more precise timescale (observed age gradient, period for the Sun to cross from one spiral arm to the next spiral arm, etc) and new fitting of theoretical predictions of arm formations (location of magnetic sectors in the Milky Way, etc), etc.

In Section 2, here we update and augment these tables (Tables 1 to 11) for different tracers with newly published arm tangents in the refereed literature. This catalog can be used for statistical studies, and is a valuable source for comparison with theoretical predictions. Some results are displayed (Figures 1, 2, 3) in Section 3. The conclusion follows in Section 4.

## 2. Selection criteria and data collection

Why build a catalog ? A catalog of similar objects can tell us more, by doing simple statistics on their properties (tracer offset x from the CO tracer, say), by testing predictions from theories on these objects (tracer ordering, say), by working up numerical experiments (maser tracer and its increasing offset from the shock lane with a time model, say), by correlations (comparing how a tracer property X changes when we observe a change in another property Y, say).

### 2.1 Small past catalogs

The catalogue of Englmaier & Gerhard (1999 – their Table 1) had 32 entries of the mean Galactic longitude of each arm tangent, covering 5 arm segments, with a total of 9 different arm tracers.

The catalogue of Vallée (2008 - Table 2) had 39 Galactic longitude scans of arm tangents, covering 6 arm segments, with a total of 9 arm tracers (HII, HI, CO, $^{13}$CO, dust 2.4μm, dust 60 μm, $^{26}$Al, 408 MHz relativistic electrons, thermal electrons); it allowed a more precise model of the spiral arms (fitted to the CO tracer) and the resulting more precise velocimetry maps: radial velocity versus distance from the Sun (Its Fig. 2) and radial velocity versus Galactic longitude (its Fig. 3).

A newer catalogue of Vallée (2014a - Table 3) had 43 entries, also with 6 arm segments and adding longitude scans from arm tangents using methanol masers. That catalogue allowed us to see, for the first time, the 'ordering' of arm tracers from the position of the CO tracer, in each arm segment (its Fig. 1), and its similarity when separating arm segments in 2 sets of alternating arms (its Fig. 3).

The catalogue of Vallée (2014b - Table1) had an appendix (Tables 3, 4, 5) totalling 107 Galactic longitude scans of arm tangents, covering 7 arm segments, adding new arm tracers (dust 240μm, dust 870 μm, FIR [CII] & [NII] lines, warm cores); it confirmed the separation of each arm tracer inside an arm at the arm tangent point, and their ordering from hot dust to cold broad diffuse CO gas (its Fig. 2), with the hot dust nearer the inside arm edge toward the Galactic Center.

A catalogue by Hou & Han (2015) had 78 Galactic longitudes, gathered from 15 tracers in 7 arm segments.

The catalogue of Vallée (2016b – Table 3) with its separate tables for different tracers (Tables 4 to 10) totalled 125 Galactic longitude scans of arm tangents, covering 8 arm segments, and some 19 different arm tracers; it depicted the mirror-image of the dust lane being closest to the Galactic Meridian (its Fig. 1) than the broad diffuse CO peak, for each spiral arm.

### 2.2 Biggest new catalog

We searched the literature for new published tangents for the spiral arms, and we verified previously published data.

Our new catalog is composed of a master table (Table 1), and associated catalogues (Tables 2 to 8), as well as scientific products (Tables 9 to 11).

Here, after searching the literature, we now present an catalog of the mean Galactic longitude (l) of arm tangents, covering 8 arm segments (Carina l ~ 283°, Crux-Centaurus l ~ 310°, Norma l ~ 328°, Perseus start l ~ 338°, Sagittarius start l ~

346°, Norma start l ~ 018°, Scutum l ~ 030°, Sagittarius l ~ 050°), using 21 different arm tracers (dust, masers, stars, gas, etc), at different wavelengths (radio, infrared, optical, etc).

**Table 1** shows the updated statistical mean Galactic longitude, for each arm tracer, and for each arm (with 50 single individual entries, and 47 entries being the means of individual entries from Tables 2 to 8). Following each tables 2 to 8, the last row gives the statistical mean and root mean square r.m.s. of the above rows (Galactic longitudes; radial velocities), with an equal weight for two entries or more. The linear separation between arm tracers is calculated, from the distance to the tangential point (given in column 5 of Table 1, at the row for $^{12}$CO at 8' beamwidth), and the angular separation from $^{12}$CO Galactic longitude (column 4 in Table 1); their product gives the linear separation.

Individual arm tracer tangents are given for $^{13}$CO (**Table 2** with 7 individual entries), $^{12}$CO (**Table 3** with 51 entries), for HII complexes (**Table 4** with 20 entries), for radio masers (**Table 5** with 36 entries), for the HI atom (**Table 6** with 19 entries), for 870μm dust (**Table 7** with 13 entries), and for 1 to 8 μm diffuse old stars and NIR star counts (**Table 8** with 9 entries).

Thus, in all these tables 1 to 8, we catalogued over two-hundreds single individual entries, nearly doubling the number in the last catalogue. Now this new catalog of 'ArmTangents' displays 205 Galactic longitude values of arm tangents, covering 8 arm segments and some 21 arm tracers, being roughly quintupling the old number of 43 entries in Vallée (2014a).

The arrival of many new individual entries has improved the means of the arm tangents. Thus comparing from Table 5 in Vallée (2016b) and from Table 3 (here) the mean tangent longitude of the Sagittarius arm in the CO tracer went from 50.5° ±0.9° to 50.7° ±0.5°. The mean longitude in the CO tracer did not change outside their errors. Some seldom used arm tracers, having less individual scans in Galactic longitude, have changed slightly.

**Table 9** shows the linear separation (from the angular separation in Table 1) of each mean longitude tracer, from the mean longitude for broad diffuse CO, at each arm tangent. As seen before (Vallée 2016b), the linear separation can be regrouped into one near 0 pc (CO, etc), one near 100 pc (electrons, etc), one near 150 pc (masers, etc), and one near 300 pc (hot dust).

**Table 10** shows the mean radial velocity of an arm tracer, at each arm tangent, from data in Table 1 (when available).

**Table 11** shows the individual arm tracers used in each spiral arm segment. As seen in this table, most arm tangents were observed in the radio regime (149 / 205 = 73%).

### 3. Results.

In this study, we presented above an updated catalog of spiral arm tangents. Some results from the catalog are as follows. A spiral arm review of the disk of the Milky Way galaxy was given in Vallée (2017b). Here we employed our global symmetric (top-down) 4-arm spiral model, based on fitting the Galactic-wide locations of the CO arm tangents in disk space (see Table 1 for the mean CO arm tangents), with a pitch angle of -13.1° for the arm, a distance of the Sun to the Galactic Center of 8.1 kpc, and each arm starting at 2.2 kpc from the Galactic Center (ignoring the small Local Arm).

A word about the Local Arm: it is not spiral, and it is not an arm – it is more like a grouping of segments. Vallée (2018a) has collated most tracers for these Local segments, showing numerous 'fingers' and 'bridges' (all imagined with scant data - see his Fig.2) – in no way can it be called 'spiral'. The peak position of each different tracer is shown to be random – in no way can it be called arm-like (masers are not aligned near a dust lane as in a long arm). Recent Gaia EDR3 data confirms the smallness of this Local arm segment: from open star clusters (Fig. 5a in Hao et al 2021); from red clump stars (Fig.3a in Lin wet al 2022); from
Giant Molecular Clouds (Fig.2 in Hou 2021); from local stars (Fig.3a in Martnez-Medina et al 2022). Their origin and formation have been discussed elsewhere, and possibly to be due to tidal encounters with a nearby dwarf galaxy (see section 2.2 in Vallée 2021b).

We employ a flat rotation curve, from a Galactic radius of 2.2 to 13 kpc (see Drimmel & Poggio 2018 - their equation 3).

More model details have been given in Vallée (2017a; 2017c). This model has been updated for the circular velocity of the local standard of rest (233 km/s) and updated for the distance from the Sun to the Galactic Center (8.1 kpc; see Abuter et al et al 2019) – the model was refitted to the CO gas tangent in Table 1 as described in Figure 3 in Vallée (2022).

The CO arm tangents are selected because of the more numerous observations, giving a better precision in the statistics done for the mean and r.m.s. of the arm tangent's longitude. A constant pitch angle was sought for all the arms, and he value was fitted and found Galactic to be 13 degrees (see Figure 4 in Vallée 2022).

**Figure 1** shows this spiral arm model, for Galactic Quadrant IV, toward the inner segment of the long Perseus arm. The observed arm tangent in CO is shown (dashed line), along with the model spiral arm (in yellow). Independently, the observed arm tangent in 2.4μm dust is shown (dashed line), and it is observed to be at an offset from the CO arm tangent; this offsetted dust tangent is closer to the direction of the Galactic Center. The orbiting gas flows (arrows) will reach the dust lane first, going clockwise around the Galactic Centre. The theoretical orbits of the gas are not exactly circular, as each theory predicts a small eccentricity (fig. 3 in Roberts 1975; Gittins & Clarke 2004, fig. 3 in Dobbs & Pringle 2010).

A similar but mirror-reversed figure was done for the spiral arms in Galactic Quadrant I (see Vallée 2021b – his fig.3).

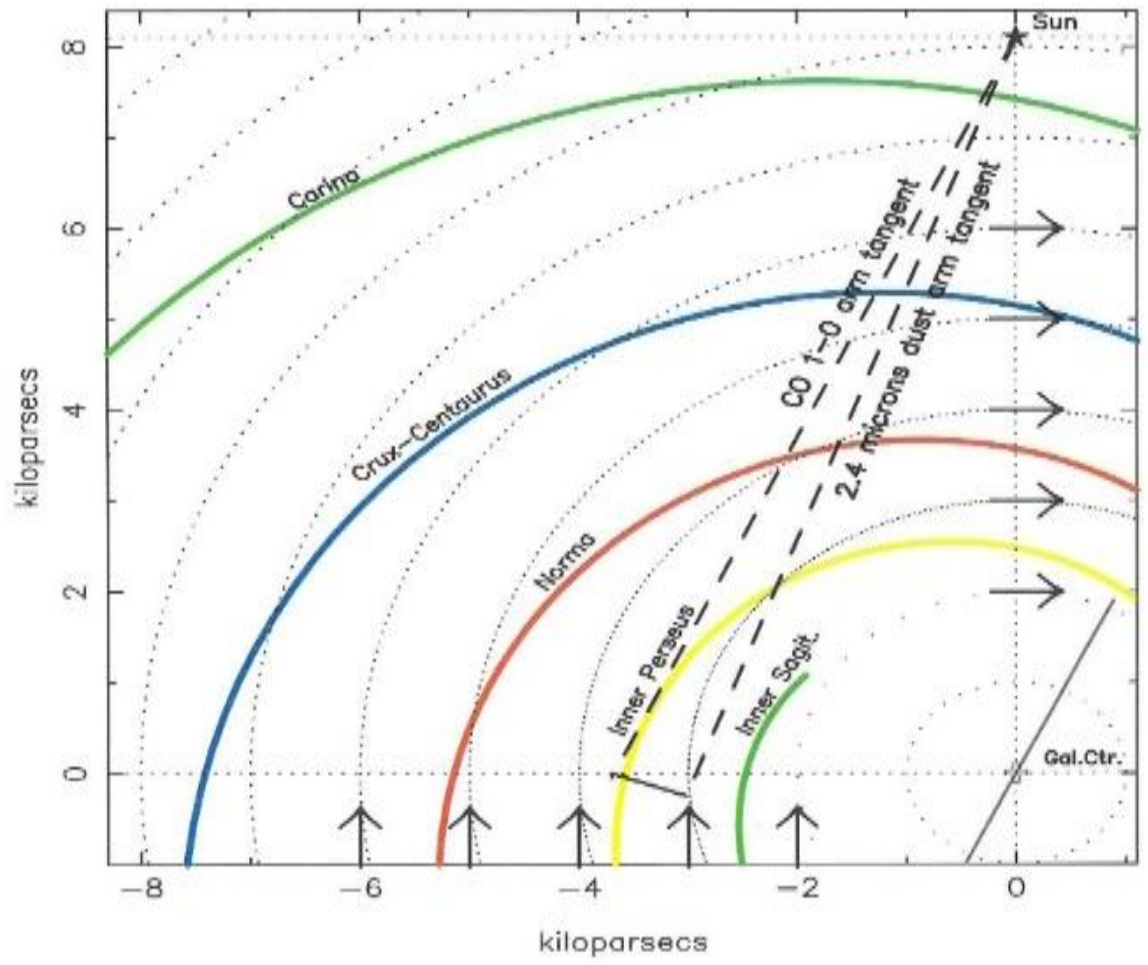

Figure 1. The mean longitude of the CO arm tangent from the Sun down to the Perseus (upper dashed line) arm differs by a few degrees from the mean longitude for the dust 2.4μm arm tangent (lower dashed line). The short black crosscut at the arm tangents near (x,y) = (-3.0, 0.0 kpc) is about 300 pc long. There, the line of sight inside the Perseus arm is shorter for the CO arm tangent (about 1.5 kpc) than for the dust arm tangent (about 3 kpc). Gas flow (black arrows, at bottom and at right) are going clockwise along a roughly circular orbit around the Galactic Center (dotted circles), entering the arm on its inner arm edge. For a similar figure in Galactic quadrant I, see Vallée (2021b – fig.3).

We can see in Table 1 that the arms in Galactic Quadrant IV (left of the Galactic Meridian) have offsets reversed from those seen in the arms in Galactic Quadrant I.

Location of arm tracers. **Figure 2** shows the arms in Galactic Quadrant IV (Carina, Crux-Centaurus, Norma, Inner Perseus), and in Galactic Quadrant I (Sagittarius, Scutum, Norma), beyond a Galactic radius of 3.1 kpc. All data are given in Table 1 above. The results for Fig.2a (Galactic quadrant IV) are the mean separation of an arm tracer for all arms in Galactic quadrant IV (from Table 1). The results for Fig.2b (Galactic quadrant IV) are the mean separation of an arm tracer for all arms in Galactic quadrant I (from Table 1).

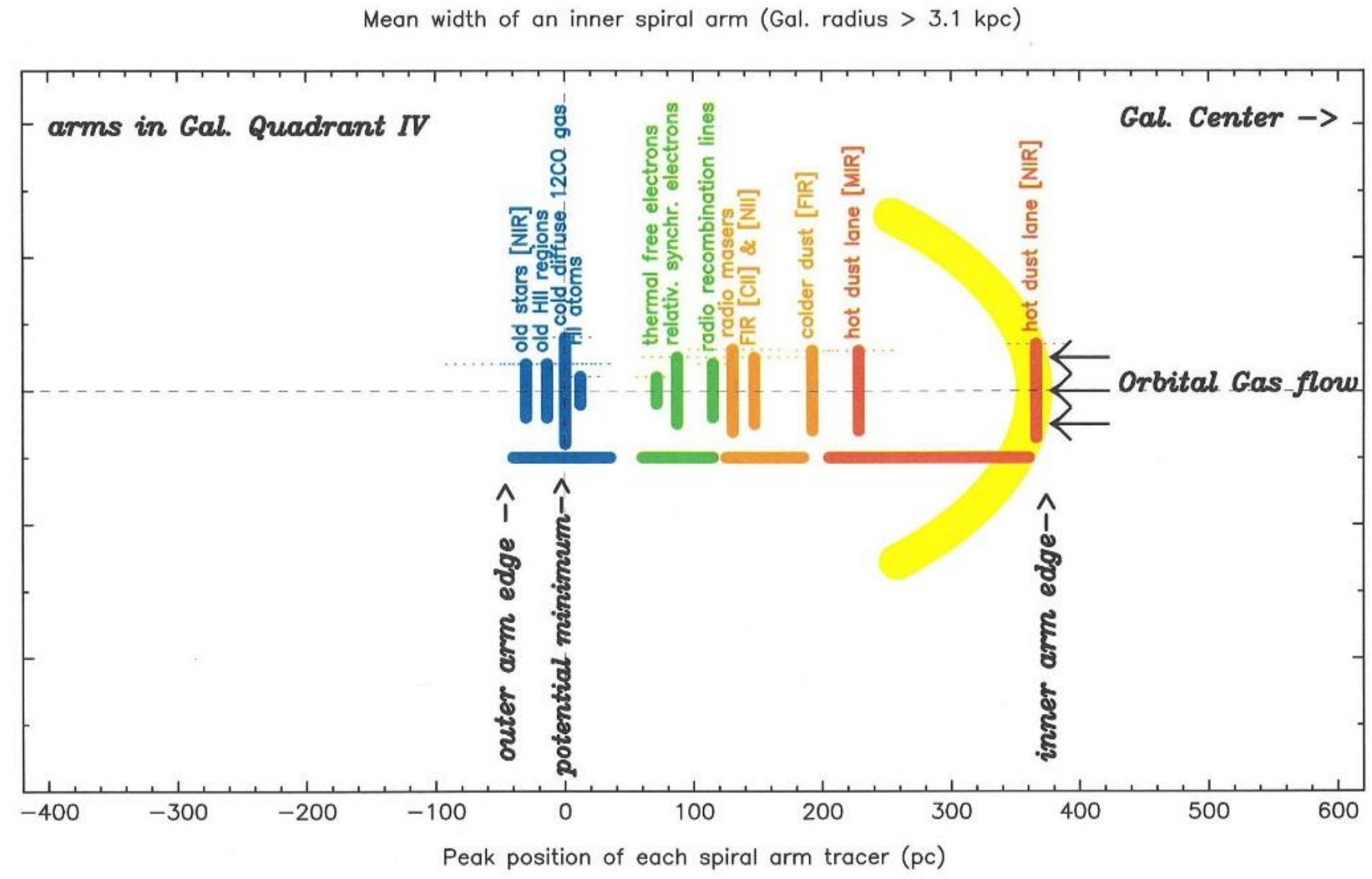

Figure 2(a). Mean offsets of arm tracers in Galactic Quadrant IV. The direction to the Galactic Center is to the right. The orbital gas flow comes from the right, and the approximate location of the compressed area is shown in yellow. The distance, from the NIR dust to the broad diffuse CO peak, is about 360 pc.

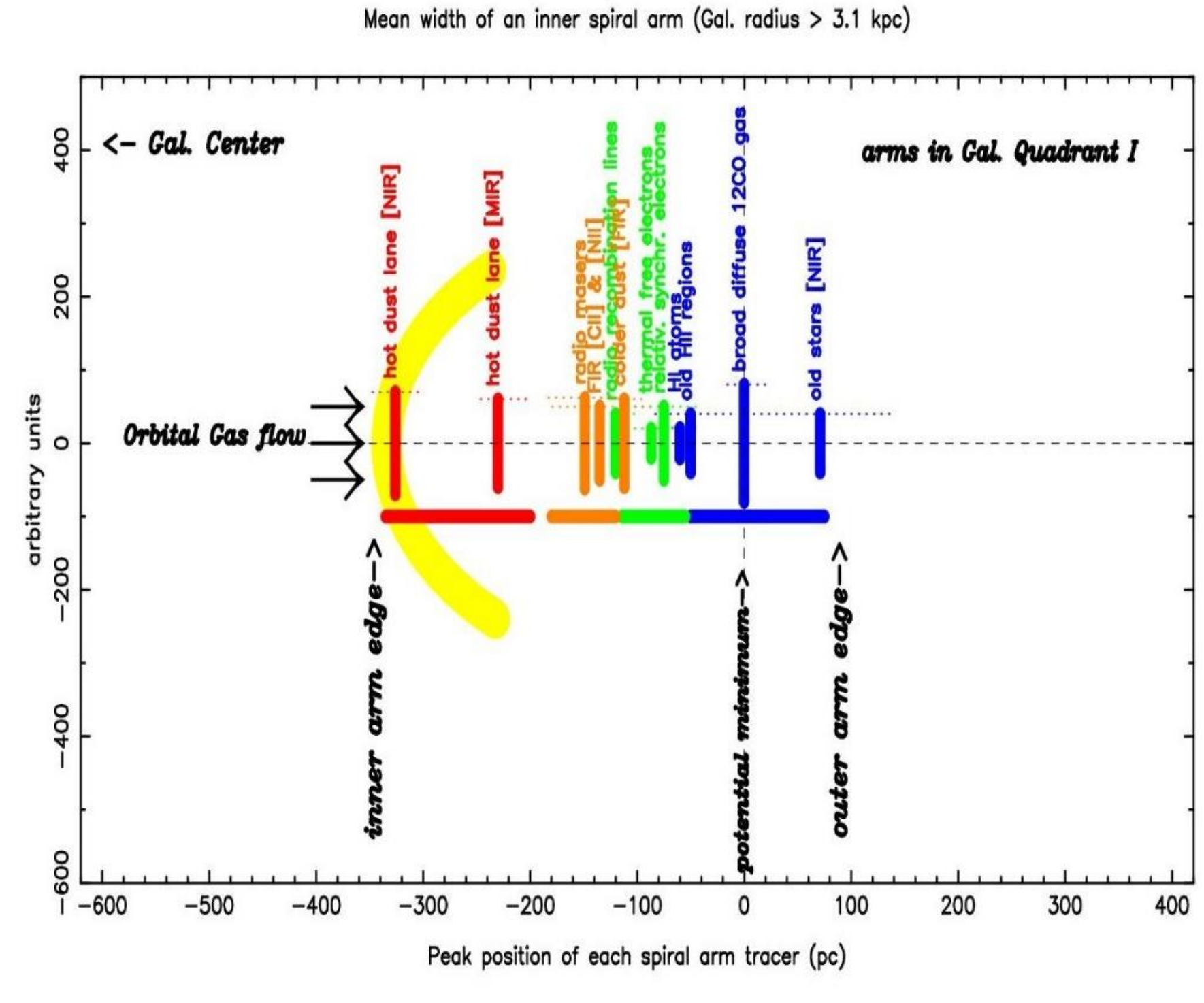

Figure 2(b). Mean offsets of arm tracers in Galactic Quadrant I. The direction to the Galactic Center is to the left. The orbital gas flow comes from the left, and the approximate location of the compressed area is shown in yellow. The distance, from the NIR dust to the broad diffuse CO peak, is about 340 pc.

This figure is made using arms in the Galactic radius range from 3.1 kpc up to 8.1 kpc (Sun's location). This implies that the co-rotations is beyond the Galactic radius of the location of the Sun.

Figure 2a and 2b constitute a confirmation of what was found 6 years ago, then having about half as many observed data. Thus our new results are a confirmation of earlier results (fig. 2a and 2b in Vallée 2016b); even the distance from the NIR dust and the broad diffuse CO peak has not changed, then in 2016 being in Galactic Quadrant IV (about 360 pc) and in Galactic Quadrants I (about 340 pc). We can be confident that if we could double again the total number of longitudinal arm tangents, the picture will not change much (or at all).

It is important to realize that a tracer may be in a different state at different locations in Figure 2 – thus the broad diffuse CO 1-0 gas (hpbw near 8') appears near the potential minimum / outer arm edge in Figure 2, while the clumpy narrow CO gas (hpbw near 50") appears near the dust lane / inner arm edge in Figure 2. Hence one should avoid talking about a 'gas arm' - instead, there is a narrow 'clumpy CO gas arm' near the shock (inner arm edge), and a broad 'diffuse CO gas arm' near the potential minimum (outer arm edge). Similarly, one should avoid talking of a homogeneous 'stellar arm' - instead, there is a 'protostellar arm' near the maser lane (inner arm edge) in Figure 2, and an 'old HII region arm and old star arm' near the potential minimum (outer arm edge).

Arm width (NIR dust to broad diffuse CO peak). We can compare the tracer offsets in both Galactic Quadrants. From Figure 2 for Galactic quadrant IV (Fig. 2a), we find a mean distance from NIR dust to broad diffuse CO peak of 360 pc, which is about 4.2% of the distance to the next spiral arm ( $0.25 \times 2 \pi R_{Gal}$, with a median $R_{Gal}$ = 5.4 kpc). From a similar figure for Galactic quadrant I (Fig. 2b), we find a mean distance from NIR dust to broad diffuse CO peak of 320 pc, which is about 3.9% of the distance to the next spiral arm ( $0.25 \times 2 \pi R_{Gal}$, with a median $R_{Gal}$ = 5.2 kpc).

In the literature, we can find the separation between NIR dust (density wave shock) and broad diffuse CO peak (density wave potential minimum), in percent of the arm separation from the next arm. This observed 3.9% to 4.2% arm width (NIR dust to broad diffuse CO) can be compared to the theoretically predicted arm width (between the compressed or shocked entrance of the orbiting gas, up to the 'potential minimum'), as compared to the arm separation (from the potential minimum up to the next potential minimum of the next arm), assuming 4 arms around a circle. Thus one reads 3.1% in Fig. 6 and 5.0% in Fig. 5 of Roberts (1969). It is read as 3.7% in Fig.11 of Gittins & Clarke (2004). It is read as 3.7% in Fig.2 in Roberts (1975). It is read as 6.4% in Fig. 2 in Tosa (1973). It is read as 7.3% in Figure 1 of Wielen (1979). Hence our observations fit nicely in the middle of the range of predictions, which averages 4.9 ±1.7 % (3.1, 5.0, 3.7, 3.7, 6.4, 7.3)

Elsewhere in 24 nearby spiral galaxies, a rough separation between a tracer near the dust lane and one near the potential minimum gave a median of 326 pc (Table 3 in Vallée 2020b) at a median Galactic radius of 4 kpc, or an arm width of 5.1% (for 4 arms).

Age gradient. With the help of published theoretical models (from the literature), we ascertain masers at 0.7 - 1.0 Myrs (orange zone), ionised radio recombination lines at 2 - 2.2 Myrs (green zone), and the diffuse broad CO peak at 3.6 - 4.0 Myrs (blue zone). Predicted ages of ultracompact HII regions are 0.7 Myrs (Xie et al 1996) and 0.4 Myrs (Wood & Churchwell 1989), while those of optically visible young compact HII regions are 2.2 Myrs (Reggiani et al 2011) and 1.5 Myrs (Hunt & Hirashita 2009).

**Figure 3** shows the results. In each colored zone, we found a model age from the literature and assumed each tracer in that colored zone to be at about the same age (see above). The dashed line is the age gradient, with a slope of **11.3 ±2 Myr/kpc** obtained by least-squares-fitting of the observational data specified just above; the fit is done with 2 variables (tracer separation from dust in pc; time since tracer birth in Myr) over 4 data (origin/red 0;0, masers/orange 190; 0.7, HII regions/green 210; 2.0, CO/blue 320;3.8).

Taking the inverse of the age gradient yields the relative speed away from the dust/shock zone; this gradient yields a relative speed of 88 ±10 pc/Myrs = **87 ±10 km/s**. This new value compares well with earlier data (Vallée 2021b had 81 ±10 km/s; Vallée 2022 had 76 ±10 km/s), within the errors, and the different methods used (linear, least-squares fit).

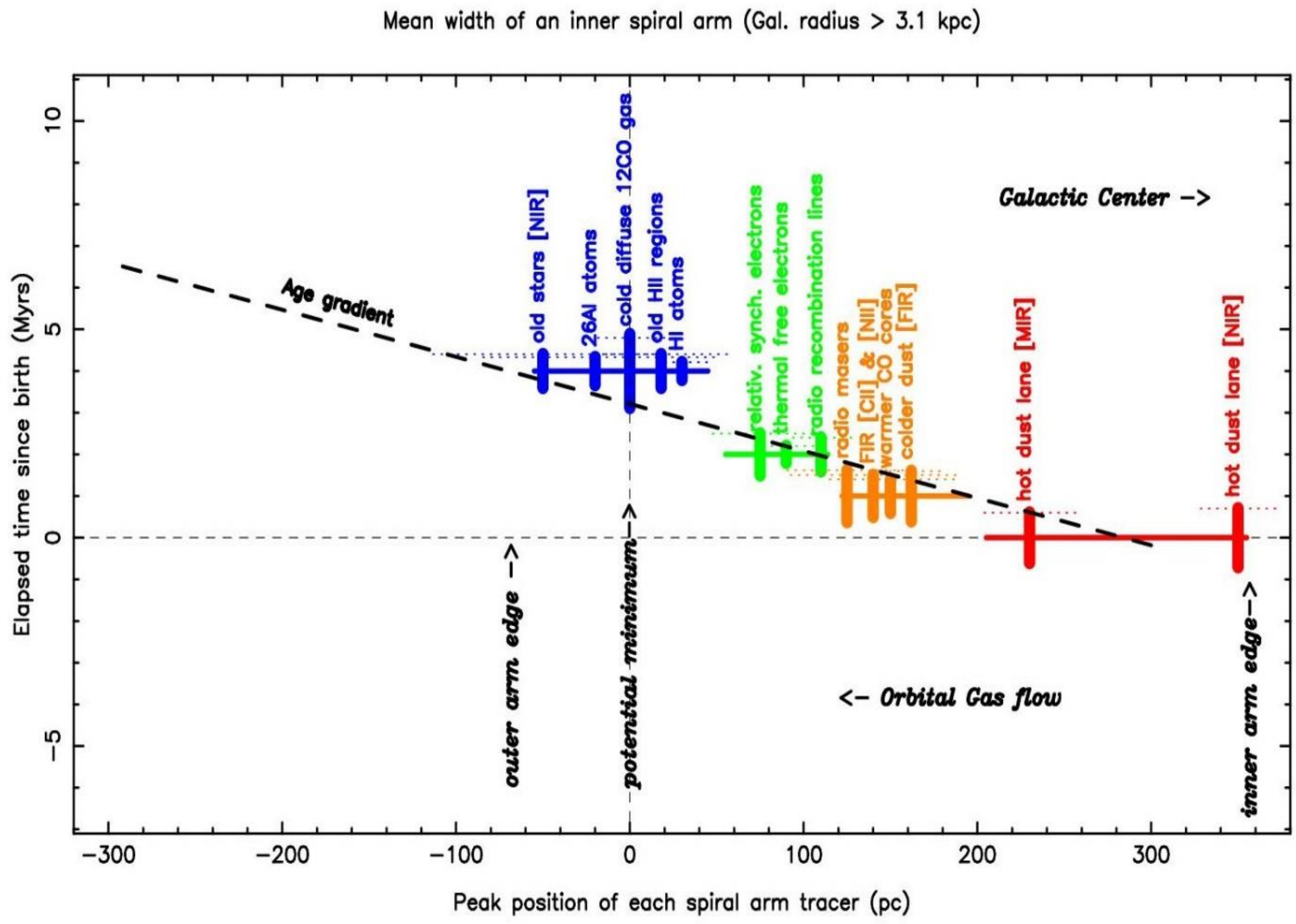

Figure 3. The distribution of each arm tracer, across a typical spiral arm (about 320 parsecs). The orbital gas flow comes from the right. The inner arm edge starts at right (red zone, compression or shock lane, hot dust), followed by the orange zone (radio masers, colder dust), then the green zone (thermal and relativistic electrons, recombination lines from HII regions), and finally the blue zone (potential minimum, broad diffuse CO gas, older stars). Individual tracer data located beyond a Galactic radius of 3.1 kpc (Tables 1 to 8) are averaged over all arms (Sagittarius-Carina, Scutum-Crux-Centaurus, Norma, Inner Perseus), to give Table 9 employed in this figure. The orbiting gas flowing around the Galactic Center in a roughly circular orbit enters the dust lane from the right, in this figure. The vertical coordinate, in Myrs, gives the elapsed time since the birth of a tracer. The black dashed line is the fitted age gradient (right to left).

In future works, it might be possible to compare/verify this calculated relative speed with astrometry measurements, such as Gaia OB stars, or other type of stars with different ages.

#### 4. Conclusion

Here we have assembled over two hundred published measurements of the Galactic longitudes of arm tangents, improving on previous smaller published tables. The master table (Table 1) with all tracers and their mean longitudes, and the individual tracer tables (Tables 2 to 8), together constitute our catalog of the tangents to the spiral arms. These tables 1 to 8 provide the largest catalog of arm tangents published so far. The mean radial velocity offset of different arm tracers (Table 10) and the contribution of all arm tracers with wavelengths (Table 11) are given.

As mentioned in Figure 1 and Table 9, there is a separation of some arm tracers among each others (CO versus dust). As mentioned in Figure 2, there is a mirror-image of this separation in Galactic Quadrants IV (shock at right) versus I (shock at left). As mentioned in Figure 3, there is an age gradient from the shock to the dust lane, within 320pc, at 11.3 Myrs/kpc, giving by inversion a relative gas speed of about 87 km/s from the dust lane.

Elsewhere, results published so far include these: a) an estimate of the angular speed of the density wave spiral pattern (shock front) – Vallée (2021a); b) an estimate and variation of the spiral arm width, from multi-tracers, with increasing Galactic radius – Vallée (2020a); c) an estimate of the co-rotation radius in the disk of the Milky Way – Vallée (2019); d) an estimate of the large-scale pitch angle of each spiral arm, using arm tangents from two Galactic quadrants – Vallée (2017a), Vallée (2015); e) an estimate of the starting point of each spiral arm, near the Galactic nucleus – Vallée (2016a); f) a location of a mirror image of arms, across the Galactic Meridian - Vallée (2016b); g) a comparison of 4 different theories for the formation of spiral arms – Vallée (2022); h) mapping the terrestrial impacts (extinctions) in time onto the passage of spiral arm structures (arm tracers) – Gillman et al (2019).

In the future, one could compare the arm tangent at the maximum intensity of the synchrotron emission and its location inside a spiral arm, as well as the radial distance of a spiral arm segment and whether there the azimuthal magnetic field is clockwise or inversed (counterclockwise). The publication of this catalog may spur other researches, and comparisons with other catalogues (Gaia, etc), so that new results may emerge in the future.

**Acknowledgements.**

Figure production made use of the PGPLOT software at the NRC Herzberg Astro & Astro Research Centre in Victoria, BC, Canada. I thank a referee for numerous clarifications.

**Declarations.**

**Funding**: the facilities of the National Research Council of Canada in Victoria BC were employed.
**Conflicts of interest:** None.
**Availability of data and material** (data transparency): All data underlying this article are available in this article, and/or will be shared upon reasonable request.
**Code availability** (software application or custom code): Basic software tools were employed, in FORTRAN.

## Table 1 – Statistical mean of the tangent longitude and velocity, for each tracer in each spiral arm [a]

| Spiral arm name | Chemical tracer | Tangent galactic[b] longitude (degree) | Angular dist. [c] to $^{12}$CO (degree) | Linear separation from $^{12}$CO [d] (pc) | Tangent radial vel. $V_{rad}$ (km/s) | References |
|---|---|---|---|---|---|---|
| Carina | $^{26}$Al | 280° | -1.8° | -125 pc | - | Chen et al (1996 – fig.1) |
| | $^{12}$CO at 8' | 281.8° | 0.0° | **0 pc, at 4.0 kpc** | -8.8 | mean in Table 3 |
| | HI atom | 281.9° | 0.1°` | 7 pc | -9 | mean in Table 6 |
| | Thermal electron | 283° | 1.2° | 84 pc | - | Taylor & Cordes (1993 – fig.4) |
| | HII complex | 283.8° | 2.0° | 140pc | - | mean in Table 4 |
| | Dust 240μm | 284° | 2.2° | 154 pc | - | Drimmel (2000 –fig.1) |
| | Dust 60μm | 284° | 2.2° | 154 pc | - | Bloemen et al (1990 – fig.5) |
| | Dust 870 μm | 284.2° | 2.4° | 168 pc | - | mean in Table 7 |
| | 1.4GHz RRL | 284.3° | 2.5° | 175 pc | - | Hou & Han (2015 – table 1) |
| | Masers | 284.8° | 3.0° | 209 pc | +10 | mean of Table 5 |
| | FIR [CII] & [NII] | 287° | 5.2° | 363 pc | - | Steiman-Cameron et al (2010 – sect. 2.1) |
| Crux-Cen-taurus | $^{13}$CO | 306.5° | -3.0° | -314pc | -35 | mean of Table 2 |
| | Old stars (1-8 μm) | 307.3° | -2.2° | -230 pc | - | mean in Table 8 |
| | FIR [CII] & [NII] | 309° | -0.5° | -52 pc | - | Steiman-Cameron et al (2010 – sect. 2.1) |
| | $NH_3$ 1-1 2' cores | 309.1° | -0.4° | -42 pc | - | Hou & Han (2015 – table 1) |
| | $^{12}$CO at 8' | 309.5° | 0.0° | **0 pc, at 6.0 kpc** | -46.6 | mean in Table 3 |
| | Masers | 309.6° | 0.1° | 10 pc | -56.5 | mean of Table 5 |
| | HII complex | 309.9° | 0.4° | 42 pc | - | mean in Table 4 |
| | Thermal electron | 310° | 0.5° | 52 pc | - | Taylor & Cordes (1993 – fig.4) |
| | $^{26}$Al | 310° | 0.5° | 52 pc | - | Chen et al (1996- fig.1) |
| | Sync. 408 MHz | 310° | 0.5° | 52 pc | - | Beuermann et al (1985 – fig.1) |
| | HI atom | 310.4° | 0.9° | 94 pc | -44 | mean in Table 6 |
| | Dust 240μm | 311° | 1.5° | 157 pc | - | Drimmell (2000 – fig. 1) |
| | Dust 60μm | 311° | 1.5° | 157 pc | - | Bloemen et al (1990 – fig.5) |
| | 1.4GHz RRL | 311.2° | 1.7° | 178 pc | - | Hou & Han (2015 – table 1) |
| | Dust 870μm | 311.4° | 1.9° | 199 pc | - | mean in Table 7 |
| Norma | $^{26}$Al | 326° | -2.4° | -314 pc | - | Chen et al (1996 – fig.1) |
| | HII complex | 326.4° | -2.0° | -262 pc | - | mean in Table 4 |
| | $^{13}$CO | 327.5° | -0.9° | -118 pc | -85 | mean of Table 2 |
| | $NH_3$ 1-1 2' cores | 327.8° | -0.6° | -78 pc | - | Hou & Han (2015 – table 1) |
| | HI atom | 327.9° | -0.5° | -65 pc | -79 | mean in Table 6 |
| | Sync. 408 MHz | 328° | -0.4° | -52 pc | - | Beuermann et al (1985 – fig.1) |
| | $^{12}$CO at 8' | 328.4° | 0.0° | **0 pc, at 7.5 kpc** | -97.6 | mean in Table 3 |
| | Thermal electron | 329° | 0.6° | 78 pc | - | Taylor & Cordes (1993 – fig.4) |
| | 1.4GHz RRL | 329.3° | 0.9° | 118 pc | - | Hou & Han (2015 – table 1) |
| | Dust 870μm | 329.6° | 1.2° | 177 pc | - | mean in Table 7 |
| | [CII] at 80" | 330° | 1.6° | 209 pc | -99 | Velusamy et al (2015 – fig.7a) |
| | Dust 60μm | 330° | 1.6° | 209 pc | - | Bloemen et al (1990 – fig.5) |
| | Masers | 330.4° | 2.0° | 262 pc | -102 | mean in Table 5 |
| | Dust 240μm | 332° | 3.6° | 471 pc | - | Drimmell (2000 – fig. 1) |
| | Dust 2.4μm | 332° | 3.6° | 471 pc | - | Hayakawa et al (1981 – fig.2a) |
| Start of Perseus | $^{26}$Al | 335° | -2.0° | -262 pc | - | Chen et al (1996 – fig.1) |
| | $^{13}$CO | 335.5° | -1.5° | -196 pc | -115 | mean of Table 2 |
| | HI atom | 336.8° | -0.2° | -26 pc | - | mean in Table 6 |
| | 1.4GHz RRL | 336.9° | -0.1° | -13 pc | - | Hou & Han (2015 – table 1) |
| | $^{12}$CO at 8' | 337.0° | 0° | **0 pc, at 7.5 kpc** | -126.7 | mean in Table 3 |
| | [CII] at 80'' | 337° | 0.0° | 0 pc | -127 | Velusamy et al (2015 – fig.8b) |
| | HII complex | 337.2° | 0.2° | 26 pc | - | mean in Table 4 |
| | Masers | 337.3° | 0.3° | 39 pc | -106.7 | mean in Table 5 |
| | Dust 870µm | 337.8° | 0.8° | 105 pc | - | mean n Table 7 |
| | FIR [CII] & [NII] | 338° | 1.0° | 131 pc | - | Steiman-Cameron et al (2010 – sect. 2.1) |
| | Old stars (1-8 µm) | 338.3° | 1.3° | 170 pc | - | mean in Table 8 |
| | $NH_3$ 1-1 2' cores | 338.4° | 1.4° | 183 pc | - | Hou & Han (2015 – table 1) |
| | Sync. 408 MHz | 339° | 2.0° | 262 pc | - | Beuermann et al (1985 – fig.1) |
| | Dust 2.4µm | 339° | 2.0° | 262 pc | - | Hayakawa et al (1981 – fig.2a) |
| | Dust 60µm | 340° | 3.0° | 393 pc | - | Bloemen et al (1990 – fig.5) |
| Start of Sagit-tarius | $^{12}$CO at 8' | 342° | 0.0° | **0 pc, at 7.5 kpc** | -136.0 | mean in Table 3 |
| | Dust 870 µm | 343° | 1.0° | 131 pc | - | mean in table 7 |
| | $^{26}$Al | 346° | 4.0° | 524 pc | - | Kretschmer et al (2013 – fig.9a) |
| | Masers | 348.0° | 6.0° | 785 pc | -120 | mean in Table 5 |
| | Old stars (1-8 µm) | 348° | 6.0° | 785 pc | - | mean in Table 8 |
| Start of Norma | Dust 870µm | 025° | -1.5° | -170 pc | - | mean of Table 7 |
| | $^{12}$CO at 8' | 023.5° | 0.0 | **0 pc, at 6.5 kpc** | +125.0 | mean of Table 3 |
| | Old stars (1-8 µm) | 019° | 4.5° | 510 pc | - | mean in Table 8 |
| | Sync. 6 GHz | 016° | 7.5° | 850 pc | - | Hayakawa et al (1981 – fig. 2b) |
| | Masers | 015.5 | 8.0° | 907 pc | +105.0 | mean in Table 5 |
| Scutum | $^{12}$CO at 8' | 032.4° | 0° | **0 pc, at 5.5 kpc** | +95.0 | mean in Table 3 |
| | $^{26}$Al | 032° | 0.4° | 38 pc | - | Chen et al (1996 – fig.1) |
| | Thermal electron | 032° | 0.4° | 38 pc | - | Taylor & Cordes (1993 – fig.4) |
| | Sync. 408 MHz | 032° | 0.4° | 38 pc | - | Beuermann et al (1985 – fig.1) |
| | $^{13}$CO | 031.8° | 0.6° | 58 pc | +95.0 | mean in Table 2 |
| | Old stars (1-8 µm) | 031.3° | 1.1° | 106 pc | - | mean in Table 8 |
| | HII complex | 031.3° | 1.1° | 106 pc | +100.0 | mean in Table 4 |
| | HI atom | 031.0° | 1.4° | 134 pc | - | mean in Table 6 |
| | Dust 240µm | 031° | 1.4° | 134 pc | - | Drimmell (2000 – fig. 1) |
| | Dust 870µm | 030.9° | 1.5° | 144 pc | - | mean in Table 7 |
| | 1.4GHz RRL | 030.8° | 1.6° | 154 pc | - | Hou & Han (2015 – table 1) |
| | Warm $^{12}$CO cores | 030° | 2.4° | 230 pc | +95 | Solomon et al (1985 – fig. 1b) |
| | [CII] at 12'' | 030° | 2.4° | 230 pc | +114 | Velusamy et al (2012 – fig.2, Table 1) |
| | FIR [CII] & [NII] | 030° | 2.4° | 230 pc | - | Steiman-Cameron et al (2010 – sect 2.1) |
| | Dust 60µm | 030° | 2.4° | 230 pc | - | Bloemen et al (1990 – fig.5) |
| | Masers | 029.6° | 2.8° | 269 pc | +100.8 | mean in Table 5 |
| | Dust 2.4µm | 029° | 3.4° | 326 pc | - | Hayakawa et al (1981 – fig. 2a) |
| Sagit-tarius | Old stars (1-8 µm) | 055.0° | -4.3° | -248 pc | - | mean in Table 8 |
| | $^{12}$CO at 8' | 050.7° | 0.0° | **0.0 pc, at 3.3 kpc** | +55.3 | mean in Table 3 |
| | HII complex | 050.7° | 0.0° | 0 pc | +61.0 | mean in Table 4 |
| | $^{13}$CO | 050.7° | 0.0° | 0 pc | +60 | mean in Table 2 |
| | HI atom | 050.6° | 0.1° | 6 pc | - | mean in Table 6 |
| | Masers | 050.2° | 0.5° | 29 pc | +65.5 | mean in Table 5 |
| | Dust 240µm | 050° | 0.7° | 40 pc | - | Drimmell (2000 – fig. 1) |
| | FIR [CII] & [NII] | 050° | 0.7° | 40 pc | - | Steiman-Cameron et al (2010 – sect. 2.1) |
| | 1.4GHz RRL | 049.2° | 1.5° | 86 pc | - | Hou & Han (2015 – table 1) |
| | Dust 870µm | 049.1° | 1.6° | 92 pc | - | mean in Table 7 |
| | Warm $^{12}$CO cores | 049° | 1.7° | 98 pc | +60 | Solomon et al (1985 – fig. 1b) |
| | Sync. 408 MHz | 049° | 1.7° | 98 pc | - | Beuermann et al (1985 – fig. 1) |
| | Excess Faraday RM | 048.2° | 2.5° | 144 pc | - | Shanahan et al (2019 – fig.3a) |

| | | | | | |
|---|---|---|---|---|---|
| Thermal electron | 048° | 2.7° | 156 pc | - | Taylor & Cordes (1993 – fig.4) |

Notes:

(a): Published since 1980. Updating Table 3 in Vallée (2016b)

(b): When there are 2 or more published reports for a given arm tracer in a given spiral arm, then a separate table is provided.

(c): Angular distance from arm center, being positive towards arm's inner edge (towards the Galactic Center), and negative in other direction (towards the Galactic anti-center).

(d): Linear separation from the arm center ($^{12}CO$), after converting the angular separation at the arm distance from the sun; using 8.1 kpc for the distance of the Sun to the Galactic Center (see Fig. 1 in Vallée 2020a).

**Table 2 - Observed tangent longitude and velocity for the $^{13}CO$ J=1-0 and 2-1 tracers [(a)(b)]**

| Arm name | Tangent Galactic longitude (degree) | Teles-cope HPBW (arc) | Survey name | Tangent radial vel. $V_{rad}$ (km/s) | Reference |
|---|---|---|---|---|---|
| Crux-Centaurus | 306.5° | 30'' | Catalog 12m J=2-1 | -35 | Schuller et al (2021 – fig.6) |
| Norma | 327.5° | 30'' | Catalog 12m J=2-1 | -85 | Schuller et al (2021 – fig.6) |
| Perseus start | 335.5° | 30'' | Catalog 12m J=2-1 | -115 | Schuller et al (2021 – fig.6) |
| Scutum | 030.5° | 46" | Galactic Ring J=1-0 | - | Hou & Han (2015- table 1) |
| | 033° | 3' | Bell Labs J=1-0 | +95 | Stark & Lee (2006 – fig.1, v= +95 km/s) |
| | 031.8° ±1.8° | mean and r.m.s. | | +95 | |
| Sagittarius | 049.4° | 46" | Galactic Ring J=1-0 | - | Hou & Han (2015- table 1) |
| | 052° | 3' | Bell Labs J=1-0 | +60 | Stark & Lee (2006 – fig.1, v= +60 km/s) |
| | 050.7° ±1.8° | mean and r.m.s. | | +60 | |

Notes:
(a): Published since 1980.
(b): updating a similar table in Vallée (2016b).

## Table 3 - Observed tangent longitude and velocity for the diffuse broad $^{12}$CO J=1-0 tracer (a) (b)

| Arm name | Tangent Galactic longitude (degree) | Teles-cope HPBW (arc) | Survey name | Tangent radial vel. $V_{rad}$ (km/s) | Reference |
|---|---|---|---|---|---|
| Carina | 280° | 8.8’ | Columbia | -10 | Alvarez et al (1990 – tables 4 and 1) |
| | 282° | 8.8’ | Columbia | -8 | Grabelsky et al (1987 – fig.4 and 6) |
| | 282° | 8.8’ | Columbia | -8 | Grabelsky et al (1988 – fig.1 and 4) |
| | 282° | 8.8’ | Columbia | -10 | Bronfman et al (2000b – tab. 2, fig.2) |
| | 282.0 | 8.8’ | Columbia | - | Hou & Han (2015- table 1) |
| | 283° | 8.8’ | Columbia | -8 | Bronfman et al (2000a – Sect. 3.4, Fig.5) |
| | 281.8° ±1.0° | mean and r.m.s. | | -8.8 ±1.1 | |
| Crux-Cen-Taurus | 308° | 8.8’ | Columbia | -48 | Bronfman et al (2000a – sect. 3.4, fig.5) |
| | 308° | 8.8’ | Columbia | - | Bronfman (2008 – sect. 4) |
| | 309° | 8.4’ | CfA | -42 | Dame & Thaddeus (2011 – fig.4, fig.2) |
| | 309° | 8.8’ | Columbia | -40 | Bronfman et al (2000b – table 2, fig.2) |
| | 309° | 8.8’ | Columbia | -44 | Bronfman et al (1988 – fig. 6) |
| | 310° | 8.8’ | Columbia | - | Alvarez et al (1990 – table 4) |
| | 310° | 8.8’ | Columbia | -50 | Bronfman et al (1989 – sect. 4, fig.8) |
| | 310° | 8.8’ | Columbia | -52 | García et al (2014 – table 3, fig.9) |
| | 311° | 8.8’ | Columbia | -50 | Grabelsky et al (1987 – fig.4 and 6) |
| | 311.0° | 8.8’ | Columbia | - | Hou & Han (2015- table 1) |
| | 309.5° ±1.1° | mean and r.m.s. | | -46.6 ±4.6 | |
| Norma | 328° | 8.8’ | Columbia | -108 | Alvarez et al (1990 – tab. 4 and 1) |
| | 328° | 8.8’ | Columbia | -90 | Bronfman et al (1988 – fig. 7) |
| | 328° | 8.8’ | Columbia | -95 | Bronfman et al (1989 – sect. 4, fig.7) |
| | 328° | 8.8’ | Columbia | - | Bronfman (1992 – fig.6) |
| | 328° | 8.8’ | Columbia | -105 | Bronfman et al (2000a – sect. 3.4, fig.5) |
| | 328° | 8.8’ | Columbia | - | Bronfman et al (2000b – table 2) |
| | 328° | 8.8’ | Columbia | - | Bronfman (2008 - sect. 4) |
| | 328.3 | 8.8’ | Columbia- | - | Hou & Han (2015- table 1) |
| | 330° | 8.8’ | Columbia | -90 | García et al (2014 – table 3, fig.2) |
| | 330° | 8.8’ | Columbia | - | Grabelsky et al (1987 – sect. 3.1.2) |
| | 328.4° ±0.8° | mean and r.m.s. | | -97.6 ±8.4 | |
| Start of Perseus | 336° | 8.8’ | Columbia | - | Bronfman et al (1989 – sect. 4) |
| | 336° | 8.8’ | Columbia | - | Bronfman (2008 – sect. 4) |
| | 336.7 | 8.8’ | Columbia | - | Hou & Han (2015- table 1) |
| | 337° | 8.8' | Columbia | -130 | Alvarez et al (1990 – tab. 1 and 4) |
| | 337° | 8.8' | Columbia | -140 | Bronfman et al (2000a – sect. 3.4, fig.5) |
| | 337° | 8.8' | Columbia | - | Bronfman et al (2000b - table 2) |
| | 338° | 8.8' | Columbia | - | Dame & Thaddeus (2008 – sect. 1) |
| | 338° | 8.8' | Columbia | -110 | García et al (2014 – tab. 2 and 3) |
| | 337.0° ±0.8° | mean and r.m.s. | | -126.7 ±15.3 | |
| Start of Sagittarius | 340° | 8.8' | Columbia | -136 | Bronfman et al (2000b – fig.4) |
| | 344° | 8.8' | Columbia | - | Grabelsky et al (1987 – fig.6) |
| | 342.0° ±02.8° | mean and r.m.s. | | -136.0 | |
| Start of Norma | 023.5 | 8.8' | Columbia | - | Dame & Thaddeus (2008 – sect.1) |
| | 023.5 | 8.4' | CfA | +125 | Reid et al (2016 –fig.7) |
| | 023.5° | mean and r.m.s. | | +125.0 | |
| Scutum | 030.5° | 8.4' | CfA | - | Hou & Han (2015- table 1) |
| | 031° | 8.4' | CfA | +105 | Dame & Thaddeus (2011 – fig.4) |
| | 031° | 7.5' | Columbia | +90 | Dame et al (1986 – fig.4) |
| | 032° | 1.1' | NRAO | +100 | Sanders et al (1985 – fig.5b, 5a) |
| | 033° | 8.4' | CfA | +96 | Reid et al (2016 – fig.7) |
| | 034° | 7.5' | Columbia | - | Cohen et al (1980 – fig.3) |
| | 035° | 1.0' | Texas | +84 | Chiar et al (1994 – Sect. 3, Tab.1) |
| | 032.4° ±1.7° | mean and r.m.s. | | +95.0 ±8.2 | |
| Sagittarius | 049.4 | 8.4' | CfA | - | Hou & Han (2015- table 1) |
| | 050 | 1.1' | NRAO | +62 | Sanders et al (1985 – fig.5b) |
| | 051° | 7.5' | Columbia | - | Cohen et al (1980 – fig.3) |
| | 051° | 7.5' | Columbia | +50 | Dame et al (1986 – fig.9 and 3) |
| | 051° | 8.8' | Columbia | - | Grabelski et al (1988 – fig.4) |
| | 052° | 8.4' | CfA | +54 | Reid et al (2016 – fig.7) |
| | 050.7° ±0.9° | mean and r.m.s. | | +55.3 ±6.1 | |

Notes:

(a): Published since 1980.

(b): updating a similar table in Vallée (2016b).

## Table 4 – Observed tangent longitude and velocity for the HII region tracer [a][b]

| Arm name | Tangent Galactic Longitude (degree) | Range | Tangent radial vel. $V_{rad}$ (km/s) | Reference |
|---|---|---|---|---|
| Carina | 283.3° | radio-IR-opt. | - | Hou & Han (2015 – table 1) |
| | 284° | optical | - | Russeil (2003 – table 6) |
| | 284° | radio | - | Downes et al (1980 – fig.4) |
| | 283.8° ±0.3° | mean and r.m.s. | - | |
| Crux-Centaurus | 309° | optical | - | Russeil (2003 – table 6) |
| | 309° | radio | - | Downes et al (1980 – fig.4) |
| | 311.7° | radio-IR-opt. | - | Hou & Han (2015 – table 1) |
| | 309.9° ±1.6° | mean and r.m.s. | - | |
| Norma | 323° | optical | - | Russeil (2003 – table 6) |
| | 328° | radio | - | Downes et al (1980 – fig.4) |
| | 328.1° | radio-IR-opt. | - | Hou & Han (2015 – table 1) |
| | 326.4° ±2.9° | mean and r.m.s. | - | |
| Start of Perseus | 337.2° | radio-IR-opt. | - | Hou & Han (2015 – table 1) |
| | 337.2° | mean | - | |
| Scutum | 030.6° | radio-IR-opt. | - | Hou & Han (2015 – table 1) |
| | 031° | optical | - | Russeil et al (2007 – fig.4) |
| | 031° | radio | +95 | Downes et al (1980 – fig.4 and 1) |
| | 032° | optical | - | Russeil (2003 – table 6) |
| | 032° | radio | +105 | Sanders et al (1985 – fig. 5a) |
| | 031.3° ±0.6° | mean and r.m.s. | +100.0 ±7.1 | |
| Sagittarius | 046° | radio | +60 | Downes et al (1980 – fig.4 and 1) |
| | 049.4° | radio-IR-opt. | - | Hou & Han (2015 – table 1) |
| | 051° | optical | - | Russeil et al (2007 – fig.4) |
| | 051° | radio | +62 | Sanders et al (1985 – fig.5a) |
| | 056° | optical | - | Russeil (2003 – table 6) |
| | 050.7° ±3.6° | mean and r.m.s. | +61.0 ±1.4 | |

Notes:
(a): Published since 1980.
(b): updating a similar table in Vallée (2016b).

## Table 5 –Observed tangent longitude and velocity for the maser tracer [(a)(b)]

| Arm name | Tangent Galactic longitude (degree) | Maser name or range | Tangent radial vel. $V_{rad}$ (km/s) | Reference |
|---|---|---|---|---|
| Carina | 284.5° | methanol | - | Hou & Han (2015 – table 1) |
| | 285° | methanol | +10 | Green et al (2012b – fig.1) |
| | 284.8° ±0.4° | mean & r.m.s. | +10 | |
| Crux-Centaurus | 306° | methanol | -55 | Green et al (2012b – fig.1) |
| | 310.5° | methanol | -58 | Green et al (2017 – fig.4) |
| | 312.2° | methanol | - | Hou & Han (2015 – table 1) |
| | 309.6° ±3.2° | mean &r.m.s. | -56.5° ±2.1° | |
| Norma | 329.3° | methanol | - | Hou & Han (2015 – table 1) |
| | 330.5° | methanol | -102 | Green et al (2017 – fig.4) |
| | 331.5° | methanol | - | Caswell et al (2011 – Sect. 4.6.2) |
| | 330.4° ±1.1° | mean & r.m.s. | -102 | |
| Start of Perseus | 336° | methanol | -115 | Green et al (2017 – fig.4) |
| | 337.0° | methanol | - | Hou & Han (2015 – table 1) |
| | 337.0 | radio kink | -115 | Reid et al (2019 – tab.2 and fig.3) |
| | 337.5° | methanol | -90 | Caswell et al (2011 – sect. 4.6.1; fig.4) |
| | 338° | methanol | - | Green et al (2011 – sect. 3.3.1) |
| | 338° | methanol | - | Green et al (2012a – Sect.2) |
| | 337.3° ±0.8° | mean & r.m.s. | -106.7 ±14.4 | |
| Start of Sagittarius | 344° | $H_2O$; methanol | - | Sanna et al (2014 – fig. 6) |
| | 348° | methanol | -120 | Green et al (2017 – fig.4) |
| | 352° | methanol | - | Green et al (2011 – sect.4) |
| | 348.0° ±4.0° | mean & r.m.s. | -120 | |
| Start of Norma | 012° | methanol | - | Green et al (2012a – sect.2) |
| | 012° | methanol | - | Green et al (2011 – sect.4) |
| | 013° | methanol | +100 | Green et al (2017 – fig.4) |
| | 025° | radio kink | +110 | Reid et al (2019 –fig.3) |
| | 015.5° ±6.4° | mean & r.m.s. | +105.0 ±7.1 | |
| Scutum | 026° | methanol | - | Green et al (2012a – Sect.2) |
| | 026° | methanol | +100 | Green et al (2011 – sect. 3.3.1) |
| | 030° | $H_2O$ & methanol | - | Sanna et al (2014 – fig.6) |
| | 030° | meth., water | - | Reid et al (2014 – fig.1) |
| | 030.8° | methanol | - | Hou & Han (2015 – table 1) |
| | 031° | meth., water | +99 | Sato et al (2014 – fig.3, and tab..3) |
| | 031° | methanol | +110 | Green et al (2017 – fig.4) |
| | 031° | meth., water | +95 | Reid et al (2019 - fig.3) |
| | 031° | radio masers | +100 | Wu et al (2019 – fig.5) |
| | 029.6° ±2.1° | mean & r.m.s. | +100.8 ±5.5 | |
| Sagittarius | 049° | methanol | +71 | Green et al (2017 – fig.4) |
| | 049.3° | methanol | - | Hou & Han (2015 – table 1) |
| | 049.6° | methanol | +68 | Pandian & Goldsmith (2007 – sect.4, fig.4) |
| | 050° | meth., water | - | Reid et al (2014 – fig.1) |
| | 051° | meth., water | +68 | Wu et al (2014 – sect. 4.2, fig.3) |
| | 052° | meth., water | +55 | Reid et al (2019 – fig.3) |
| | 050.2° ±1 .1° | mean & r.m.s. | +65.5 ±7.1 | |

Notes:

(a): Published since 1980

(b): updating a similar table in Vallée (2016b).

## Table 6 - Observed tangent longitude and velocity for the HI atom tracer[(a)(c)]

| Arm name | Tangent Galactic Longitude (degree) | Teles-cope HPBW (arc) | Survey name | Tangent radial vel. $V_{rad}$ (km/s) | Reference |
|---|---|---|---|---|---|
| Carina | 281.2° | 36’ | Leiden | - | Nakanishi & Sofue (2016- fig.7) [(b)] |
| | 281.5° | 48’ | Parkes 18m | -9 | Grabelswky et al (1987 – fig.11 |
| | 283.0° | 36’ | LAB | - | Hou & Han (2015 – Table 1) |
| | 281.9° ±1.0° | mean and r.m.s. | | -9 | |
| Crux-Centaurus | 309.3° | 36’ | Leiden | - | Nakanishi & Sofue (2016- fig.7) [(b)] |
| | 310° | 15’ – 36’ | Hat Creek;Parkes | - | Englmaier & Gerhard (1999- table 1) |
| | 310.4° | 36’ | LAB | - | Hou & Han (2015 – Table 1) |
| | 312.0° | 48’ | Parkes 18m | -44 | Grabelsky et al (1987 – fig.11) |
| | 310.4° ±1.1° | mean and r.m.s. | | -44 | |
| Norma | 327.5° | 48’ | Parkes 18m | -79 | Grabelsky et al (1987 – fig.11) |
| | 328.0° | 36’ | LAB | - | Hou & Han (2015 – Table 1) |
| | 328.0° | 15’ – 36’ | Hat Creek;Parkes | - | Englmaier & Gerhard (1999- table 1) |
| | 328.4° | 36’ | Leiden | - | Nakanishi & Sofue (2016- fig.7)[(b)] |
| | 327.9° ±0.4° | mean and r.m.s. | | -79 | |
| Start of Perseus | 336.8° | 36’ | LAB | - | Hou & Han (2015 – Table 1) |
| | 336.9° | 36’ | Leiden | - | Nakanishi & Sofue (2016- fig.7)[(b)] |
| | 336.8° ±0.1° | mean and r.m.s. | | - | |
| Scutum | 029° | 15’ – 36’ | Hat Creek;Parkes | - | Englmaier & Gerhard (1999 – table 1) |
| | 030.8° | 36’ | LAB | - | Hou & Han (2015 – Table 1) |
| | 033.2° | 36’ | Leiden | - | Nakanishi & Sofue (2016- fig.7)[(b)] |
| | 031.0° ±2.1° | mean and r.m.s. | | - | |
| Sagittarius | 050° | 15’ – 36’ | Hat Creek;Parkes | - | Englmaier & Gerhard (1999 – table 1) |
| | 050.8° | 36’ | LAB | - | Hou & Han (2015 – Table 1) |
| | 051.0° | 36’ | Leiden | - | Nakanishi & Sofue (2016- fig.7)[(b)] |
| | 050.6° ±0.5° | mean and r.m.s. | | - | |

Notes:

(a): Published since 1980.

(b): Nakanishi & Sofue (2016) reassessed the published HI catalogs from Hartmann & Burton (1997) and Bajaja et al (2005); they added the CO survey of Dame et al (2001).

(c): updating a similar table in Vallée (2016b).

## Table 7 - Observed tangent longitude and velocity for the dust 870μm tracer [(a)] [(b)]

| Arm name | Tangent Galactic longitude (degree) | Teles-cope HPBW (arc) | Survey name | Tangent radial vel. $V_{rad}$ (km/s) | Reference |
|---|---|---|---|---|---|
| Carina | 284.2.0° | 19" | Atlasgal | - | Hou & Han (2015 – Table 1) |
| | 284.2° | mean | | - | |
| Crux-Centaurus | 311° | 19" | Atlasgal | - | Beuther et al (2012 – fig.2) |
| | 311.7° | 19" | Atlasgal | - | Hou & Han (2015 – Table 1) |
| | 311.4° ±0.5° | mean and r.m.s. | | - | |
| Norma | 327.2° | 19" | Atlasgal | - | Hou & Han (2015 – Table 1) |
| | 332° | 19 " | Atlasgal | - | Beuther et al (2012 – fig.3) |
| | 329.6° ±3.4° | mean and r.m.s. | | - | |
| Start of Perseus | 337.5° | 19" | Atlasgal | - | Hou & Han (2015 – Table 1) |
| | 338° | 19" | Atlasgal | - | Beuther et al (2012 – fig.3) |
| | 337.8° ±0.3° | mean and r.m.s. | | - | |
| Start of Sagittarius | 343° | 19" | Atlasgal | - | Beuther et al (2012 – fig.3) |
| | 343° | mean | | - | |
| Start of Norma | 025° | 19'' | Atlasgal | - | Beuther et al (2012 – fig.3) |
| | 025° | mean | | - | |
| Scutum | 030.7° | 19" | Atlasgal | - | Hou & Han (2015 – Table 1) |
| | 031° | 19" | Atlasgal | - | Beuther et al (2012 – fig.3) |
| | 030.9° ±0.2° | mean and r.m.s. | | - | |
| Sagittarius | 049° | 19" | Atlasgal | - | Beuther et al (2012 – fig.3) |
| | 049.2° | 19" | Atlasgal | - | Hou & Han (2015 – Table 1) |
| | 049.1° ±0.2° | mean and r.m.s. | | - | |

Notes:
(a): Published since 1980. (b): updating a table in Vallée (2016b).

**Table 8 - Observed tangent longitude and velocity for the 1 to 8 µm diffuse old stars (NIR star counts) tracer [(a)] [(b)]**

| Arm name | Tangent Galactic longitude (degree) | Teles-cope HPBW (arc) | Survey name | Tangent radial vel. $V_{rad}$ (km/s) | Reference |
|---|---|---|---|---|---|
| Crux-Centaurus | 307° | 21' | COBE K-band 2µm | - | Drimmel (2000 – Sect.3) |
| | 307.5° | 1.2;2" | GLIMPSE; 2MASS | - | Hou & Han (2015 – Table 1) |
| | 307.3° ±0.4° | mean and r.m.s. | | - | |
| Start of Perseus | 338.3° | 1.2";2" | GLIMPSE; 2MASS | - | Hou & Han (2015 – Table 1) |
| | 338.3° | mean | | - | |
| Start of Sagittarius | 348° | 21' | J,H,K COBE 2µm | - | Benjamin (2008 – Fig.2) |
| | 348° | mean | | - | |
| Start of Norma | 019° | 21' | J,H,K COBE 2µm | - | Benjamin (2008 – Fig.2) |
| | 019° | 1.2;2" | 4.5µm GLIMPSE | - | Benjamin (2008 – Fig.2) |
| | 019° | mean | | - | |
| Scutum | 030° | 21' | COBE K-band 2µm | - | Drimmel (2000 – Sect.3) |
| | 032.6° | 1.2";2" | GLIMPSE; 2MASS | - | Hou & Han (2015 – Table 1) |
| | 031.3° ±1.8° | mean and r.m.s. | | - | |
| Sagittarius | 055.0° | 1.2";2" | GLIMPSE; 2MASS | - | Hou & Han (2015 – Table 1) |
| | 055.0° | mean | | - | |

Notes:
(a): published since 1980.
(b): updating a similar table in Vallée (2016b).

**Table 9 – Mean linear separation (S) of each arm tracer from diffused broad $^{12}CO$, at each arm tangent [(a) (g)]**

| Mean tangent Longit. | 283° | 310° | 328° | 338°` | 346° | 018° | 030° | 050° | - |
|---|---|---|---|---|---|---|---|---|---|
| At Gal. radius (kpc) | 8.0 | 6.3 | 4.5 | 3.2 | 2.5 | 2.8 | 4.2 | 6.3 | - |
| Chemical Tracer | S in Carina arm | S in Crux-Centaurus arm | S in Norma arm | S in Start of Perseus arm | S in Start of Sagittarius arm | S in Start of Norma arm | S in Scutum arm | S in Sagittarius arm | Mean separation ±s.d.m.[(b)] |
| | (pc) | (pc) | (pc) | (pc) | (pc) | (pc) | (pc) | (pc) | (pc) |
| blue group: | | | | | | | | | |
| $^{12}CO$ at 8' | 0 | 0 | 0 | 0 | 0 | 0 | 0 | 0 | 0 ±- [(c)] |
| $^{13}CO$ | - | -314 | -118 | -196 | - | - | 58 | 0 | -114 ±67 |
| Old stars (NIR) | - | -230 | - | 170 | 785 | 510 | 106 | -248 | -50 ±110 [(f)] |
| $^{26}Al$ | -125 | 52 | -314 | -262 | 524 | - | 38 | - | -14 ±124 |
| HII complex | 140 | 42 | -262 | 26 | - | - | 106 | -7 | 8 ±58 |
| $NH_3$ 1-1 cores | - | -42 | -78 | 183 | - | - | - | - | 21 ±81 |
| HI atom | 7 | 94 | -65 | -26 | - | - | 134 | 6 | 25 ±6 |
| [CII] at 80" | - | - | -26 | 0 | - | - | 230 | - | 68 ±81 |
| green group: | | | | | | | | | |
| Thermal electron | 84 | 52 | 78 | - | - | - | 38 | 156 | 82 ±21 |
| Synch. 408 MHz | - | 52 | -52 | 262 | - | - | 38 | 98 | 80 ±52 |
| 1.4GHz RRL | 175 | 178 | 118 | -13 | - | - | 154 | 86 | 116 ±5 |
| orange group: | | | | | | | | | |
| Masers | 209 | 10 | 262 | 39 | 785 | 907 | 269 | 29 | 136 ±50 [(f)] |
| FIR [CII] & [NII] | 363 | -52 | - | 131 | - | - | 230 | 40 | 142 ±12 |
| Cold dust 870μm | 168 | 199 | 177 | 105 | 131 | -170 | 144 | 92 | 148 ±11 [(f)] |
| Warm $^{12}CO$ cores | - | - | - | - | - | - | 220 | 98 | 159 ±61 |
| Cold dust 240μm | 154 | 157 | 471 | - | - | - | 134 | 40 | 191 ±13 |
| red group: | | | | | | | | | |
| Hot dust 60μm | 154 | 157 | 209 | 393 | - | - | 230 | - | 229 ±41 |
| Hot dust 2.4μm | - | - | 471 | 262 | - | - | 326 | - | 353 ±62 |
| Combined cold dust [(d)] | 161 | 178 | 324 | 105 | 131 | -170 | 135 | 66 | 162 ±36 [(f)] |
| Combined hot dust [(e)] | 154 | 157 | 340 | 328 | - | - | 278 | - | 251 ±41 |

Notes:

(a): All data from Table 1 here.

(b): The s.d.m.(standard deviation of the mean) is from the external scatter.

(c): There is a median internal scatter of 40 pc, from the CO data in Table 3 (1.0° rms, sdm 0.3°, at 7 kpc).

(d): Statistical means made on cold dust tracers at 240μm and 870μm.
(e): Statistical means made on hot dust tracers at 2.4μm and 60μm.
(f): Excluding the two tangent arms, nearest the Galactic Meridian (Start of Sagittarius, Start of Norma).
(g): updating a similar table in Vallée (2016b).

**Table 10 – Mean radial velocity of each arm tracer, at each arm tangent (a)**

| Mean tangent Longit. / At Gal. radius (kpc) | 283° / 8.0 | 310° / 6.3 | 328° / 4.5 | 338°` / 3.2 | 346° / 2.5 | 018° / 2.8 | 030° / 4.2 | 050° / 6.3 |
|---|---|---|---|---|---|---|---|---|
| Chemical Tracer | $V_{rad}$ in Carina arm (km/s) | $V_{rad}$ in Crux-Centaurus arm (km/s) | $V_{rad}$ in Norma arm (km/s) | $V_{rad}$ in Start of Perseus arm (km/s) | $V_{rad}$ in Start of Sagittarius arm (km/s) | $V_{rad}$ in Start of Norma arm (km/s) | $V_{rad}$ in Scutum arm (km/s) | $V_{rad}$ in Sagittarius arm (km/s) |
| blue group: | | | | | | | | |
| $^{12}$CO at 8' | -8.8 | -46.6 | -97.6 | -126.7 | -136 | +125 | +95.0 | +55.3 |
| [CII] at 80" | - | - | -106 | -127 | - | - | +117 | - |
| HI atom | -9 | -44 | -79 | - | - | - | - | - |
| HII complex | - | - | - | - | - | - | +100.0 | +61 |
| $^{13}$CO | - | -35 | -85 | -115 | - | - | +95.0 | +60 |
| orange group: | | | | | | | | |
| Warm $^{12}$CO cores | - | - | - | - | - | - | +95 | +60 |
| Masers | +10 | -56.5 | -102 | -106.7 | -120 | +105 | +100.8 | +65.5 |
| Mean radial velocity: | -3 ±6 | -44 ±5 | -94 ±5 | -119 ±5 | -128 ±5 | +115 ±10 | +100 ±4 | +60 ±3 |

Notes:
(a): All data from Table 1 here.

## Table 11 – All arm tracers across the wavelength range – contribution to each arm [(a)]

| Chemical Tracer | Wave-length | No. in Carina arm | No. in Crux-Centaurus arm | No. in Norma arm | No. in Start of Perseus arm | No. in Start of Sagit-tarius | No. in Start of Norma arm | No. in Scutum arm | No. in Sagit-tarius arm | Sum of No. |
|---|---|---|---|---|---|---|---|---|---|---|
| Mean tangent Longit. | | 283° | 310° | 328° | 338°` | 346° | 018° | 030° | 050° | - |
| At Gal. radius (kpc) | | 8.0 | 6.3 | 4.5 | 3.2 | 2.5 | 2.8 | 4.2 | 6.3 | - |
| **Gamma rays:** | | | | | | | | | | |
| $^{26}$Al | 0.7μμm | 1 | 1 | 1 | 1 | 1 | 0 | 1 | 0 | 6 |
| **Optical régime:** | | | | | | | | | | |
| HII complex | 0.5μm | 1 | 1 | 1 | 0 | 0 | 0 | 2 | 2 | 7 |
| **Infrared régime:** | | | | | | | | | | |
| Old stars (NIR) | 1-8μm | 0 | 2 | 0 | 1 | 1 | 2 | 2 | 1 | 9 |
| Dust | 2.4μm | 0 | 0 | 1 | 1 | 0 | 0 | 1 | 0 | 3 |
| Dust | 60μm | 1 | 1 | 1 | 1 | 0 | 0 | 1 | 0 | 5 |
| [CII] at 80" | 158μm | 0 | 0 | 1 | 1 | 0 | 0 | 1 | 0 | 3 |
| FIR [CII]&[NII] | 158-205μm | 1 | 1 | 0 | 1 | 0 | 0 | 1 | 1 | 5 |
| Dust | 240μm | 1 | 1 | 1 | 0 | 0 | 0 | 1 | 1 | 5 |
| Dust | 870μm | 1 | 2 | 2 | 2 | 1 | 1 | 2 | 2 | 13 |
| **Radio régime:** | | | | | | | | | | |
| $^{12}$CO at 8' | 3mm | 6 | 10 | 10 | 8 | 2 | 2 | 7 | 6 | 51 |
| $^{13}$CO | 3mm | 0 | 1 | 1 | 1 | 0 | 0 | 2 | 2 | 7 |
| Warm $^{12}$CO cores | 3mm | 0 | 0 | 0 | 0 | 0 | 0 | 1 | 1 | 2 |
| $NH_3$ 1-1 cores | 1.2cm | 0 | 1 | 1 | 1 | 0 | 0 | 0 | 0 | 3 |
| Masers | 4cm | 2 | 3 | 3 | 6 | 3 | 4 | 9 | 6 | 36 |
| Synch 6 GHz | 5cm | 0 | 0 | 0 | 0 | 0 | 1 | 0 | 0 | 1 |
| HII complex | 6cm | 2 | 2 | 2 | 1 | 0 | 0 | 3 | 3 | 13 |
| Thermal electron | 6cm | 1 | 1 | 1 | 0 | 0 | 0 | 1 | 1 | 5 |
| Faraday RM | 15cm | 0 | 0 | 0 | 0 | 0 | 0 | 0 | 1 | 1 |
| HI atom | 21cm | 3 | 4 | 4 | 2 | 0 | 0 | 3 | 3 | 19 |
| 1.4GHz RRL | 21cm | 1 | 1 | 1 | 1 | 0 | 0 | 1 | 1 | 6 |
| Synch. 408 MHz | 74cm | 0 | 1 | 1 | 1 | 0 | 0 | 1 | 1 | 5 |
| Grand total: | - | 21 | 33 | 32 | 29 | 8 | 10 | 40 | 32 | 205 |

Notes:
(a): All data from Tables 1 to 8 here.